\documentclass[11pt]{article}
\usepackage{amssymb}

\newcommand{\bea}{\begin{eqnarray*}}
\newcommand{\eea}{\end{eqnarray*}}
\newcommand{\alfa}{\alpha}
\newcommand{\w}{\omega}

\newtheorem{teorema}{Theorem}
\newtheorem{lemma}{Lemma}
\newtheorem{opmerking}{Remark}
\newtheorem{definisie}{Definition}

\begin{document}

\title{The decomposition of an arbitrary $2^w \times 2^w$ unitary matrix
       into signed permutation matrices}
\author{Alexis De Vos$^1$ and Stijn De Baerdemacker$^2$ \\[2mm]
        ($^1$) Universiteit Gent, Belgium \\
        ($^2$) University of New Brunswick, Canada}

\maketitle


\begin{abstract}
Birkhoff's theorem tells that any doubly stochastic matrix can be
decomposed as a weighted sum of permutation matrices. 
A similar theorem reveals that any unitary matrix can be
decomposed as a weighted sum of complex permutation matrices.
Unitary matrices of dimension equal to a power of~2 (say $2^w$)
deserve special attention, as they represent quantum qubit circuits.
We investigate which subgroup of the signed permutation matrices
suffices to decompose an arbitrary such matrix.
It turns out to be a matrix group isomorphic to 
the extraspecial group {\bf E}$_{2^{2w+1}}^+$ of order $2^{2w+1}$.
An associated projective group of order $2^{2w}$ equally suffices.
\end{abstract}

\section{Introduction}

Let $D$ be an arbitrary $n \times n$ doubly stochastic matrix.
This means that all entries $D_{j,k}$ are real and satisfy $0 \le D_{j,k} \le 1$
and that all line sums (i.e.\ the $n$~row sums and the $n$~column sums) 
are equal to~1. 
Let P($n$) be the group of all $n \times n$ permutation matrices.
Birkhoff \cite{birkhoff} has demonstrated 
\begin{teorema}
Any $n \times n$ doubly stochastic matrix $D$ can be written 
       \[
       D = \sum_j c_j P_j
       \]
       with all $P_j \in$ P($n$) and the weights $c_j$ real, 
       satisfying both $0 \le c_j \le 1$
       and $\sum_j c_j = 1$. 
\end{teorema}
Because unitary matrices describe quantum circuits \cite{nielsen}
and permutation matrices describe classical reversible circuits \cite{devos},
the question arises whether a similar theorem holds
for matrices from the unitary group~U($n$).

It is clear that an arbitrary             $n \times n$ unitary     matrix
cannot be decomposed as a weighted sum of $n \times n$ permutation matrices. 
Indeed, such a sum always results in an   $n \times n$             matrix 
with $2n$ identical line sums.
We have shown in previous work \cite{debaerdemacker}
that a unitary matrix with the additional feature of equal linesums
can be Birkhoff decomposed as a weighted sum of permutation matrices.
However, if we loosen the requirement of a decomposition in strictly
permutation matrices, we can lift the restriction on the equal linesum
of the unitary matrix.
In \cite{prime} \cite{lausanne} it is demonstrated that an arbitrary U($n$) matrix
can be decomposed as a weighted sum of complex permutation matrices
and, in partcular, of signed permutation matrices if $n$ is equal to a power of~2, say~$2^w$.
Because it was demonstrated before by us
\cite{prime-power} that prime-powers hold interesting properties,
in the present paper, we will focus on the special case
of $n=2^w$.

The $2^w \times 2^w$ signed permutation matrices form a finite group
of order $(2^w)!\, 2^{\, 2^w}$. 
In the present paper we investigate a particular subgroup of this group, 
such that the members of the subgroup suffice to decompose
an arbitrary matrix from U($2^w$).
The construction of the subgroup in question involves
the dihedral group of order~8, 
which will be discussed in detail in the next section. 

Before investigating the dihedral group,
we make a preliminary remark about matrices:
\begin{opmerking}
We number rows and columns of any $2^w \times 2^w$ matrix 
from~$0$ to $2^w-1$
(instead of the conventional numbering from~$1$ to~$2^w$)
and each such number we represent by the $w \times 1$ matrix
consisting of the $w$~bits of the binary notation of the row-or-column number.  
\end{opmerking}
E.g.\ the upper-left entry of an $8 \times 8$ matrix~$A$ is entry 
$A_{0,0} = A_{(0, 0, 0)^T,\, (0, 0, 0)^T}$, whereas its lower-right entry is denoted
$A_{7,7} = A_{(1, 1, 1)^T,\, (1, 1, 1)^T}$.

\section{The dihedral group {\bf D}}

Unitary $2^w \times 2^w$ matrices are interpreted as quantum circuits acting on $w$~qubits.
The number $w$ is called either the circuit width or the qubit count.
For $w=1$, the single-qubit circuit is called a gate, represented by a matrix from U(2).
Below, two gates will be used as building block: the {\tt X}~gate and the {\tt Z}~gate.

The {\tt X} gate, a.k.a.\ the {\tt NOT} gate, is a classical gate,
represented by the matrix ${\tiny \left( \begin{array}{cc} & 1 \\ 1 & \end{array} \right)}$. In contrast, 
the {\tt Z} gate is a truly quantum gate,
represented by the matrix ${\tiny \left( \begin{array}{cc} 1 & \\ & -1 \end{array} \right)}$.
Together, the {\tt X} gate and the {\tt Z} gate generate a group
of order~8, consisting of the eight $2 \times 2$ matrices
\bea
M_0 = & \left( \begin{array}{cc}  1 &    \\    &  1 \end{array} \right) & = {\tt X^2} =  {\tt Z^2}  =   {\tt I} \\ 
M_1 = & \left( \begin{array}{cc} -1 &    \\    & -1 \end{array} \right) & = {\tt XZXZ} = {\tt ZXZX} = - {\tt I} \\
M_2 = & \left( \begin{array}{cc}    &  1 \\  1 &    \end{array} \right) & = {\tt X}   \\ 
M_3 = & \left( \begin{array}{cc}    & -1 \\ -1 &    \end{array} \right) & = {\tt ZXZ} = - {\tt X} \\
M_4 = & \left( \begin{array}{cc}  1 &    \\    & -1 \end{array} \right) & = {\tt Z}   \\
M_5 = & \left( \begin{array}{cc} -1 &    \\    &  1 \end{array} \right) & = {\tt XZX} = - {\tt Z} \\
M_6 = & \left( \begin{array}{cc}    &  1 \\ -1 &    \end{array} \right) & = {\tt ZX}  \\
M_7 = & \left( \begin{array}{cc}    & -1 \\  1 &    \end{array} \right) & = {\tt XZ}  \ ,
\eea
where {\tt I} is the identity gate. 
The group is isomorphic to the dihedral group {\bf D} of order~8. 
The above ordering of the indices~$j$ of the $M_j$ matrices will be elucidated later on
(see Section~3).

We note that the gates {\tt X} and {\tt Z}, completed with $i{\tt X}{\tt Z} = {\tt Y}$,
are called the Pauli matrices.
The matrix set $\{ M_0, M_6, M_1, M_7 \}$ forms a cyclic subgroup of~{\bf D}.
The matrix set $\{ M_0, M_2, M_4, M_6 \}$ does not form a            group; it
                                          does     form a projective group.
The decomposition properties of
both this projective group and the group {\bf D} itself have been
studied by Allouche et al.\ \cite{allouche}.  

The above matrices $M_j$ (with $0 \le j \le 7$) constitute a group representation
of the group {\bf D} which is irreducible. Indeed,
the matrix $M_0$ (i.e.\    {\tt I}) has trace equal to~2, 
the matrix $M_1$ (i.e.\ $-${\tt I}) has trace $-2$,
whereas the remaining six matrices are traceless.
Thus
\[
\sum_{j=0}^7 |\mbox{Tr}(M_j)|^2 = |\, 2\, |^2 + |\, -2\, |^2 = 8 \ . 
\]
Hence, the irreducibility criterion
\[
\sum_{M_j \in {\bf D}} |\mbox{Tr}(M_j)|^2 = \mbox{Order}({\bf D}) 
\]
is fulfilled.
We denote this first irrep by $R^{(1)}_j$ (with $0 \le j \le 7$).
According to Burrow \cite{burrow}, the group {\bf D} has,
besides this 2-dimensional irreducible representation,
four 1-dimensional irreps.
We denote these by $R^{(2)}_j$, $R^{(3)}_j$, $R^{(4)}_j$, and $R^{(5)}_j$.
We have $R^{(2)}_j=1$, whereas 
$R^{(3)}_j$, $R^{(4)}_j$, and $R^{(5)}_j$ equal $\pm 1$,
the character table of {\bf D} looking like
\[
\begin{tabular}{c|cccccc}
& $\{ M_0 \}$ & $\{ M_1 \}$ & $\{ M_2, M_3 \}$ & $\{ M_4, M_5 \}$ & $\{ M_6, M_7 \}$ & \\
&&&&&& \\[-3mm]  
\hline
&&&&&& \\[-3mm]  
$R^{(2)}$ & 1 &   1  &   1  &   1  &   1  &   \\
$R^{(3)}$ & 1 &   1  & $-1$ &   1  & $-1$ &   \\
$R^{(4)}$ & 1 &   1  &   1  & $-1$ & $-1$ &   \\
$R^{(5)}$ & 1 &   1  & $-1$ & $-1$ &   1  &   \\
$R^{(1)}$ & 2 & $-2$ &   0  &   0  &   0  & . \\
\end{tabular}
\]

\begin{teorema}
Any U(2) matrix $U$, i.e.\ any matrix representing a single-qubit gate,
can be written 
       \[
       U = \sum_j c_j M_j
       \]
       with all $M_j \in$ {\bf D} and the weights $c_j$ complex numbers,
       such that both $\sum c_j = 1$ and $\sum |c_j|^2 = 1$.
\end{teorema}

In order to find the values of the eight coefficients~$c_j$, 
it suffices to solve a matrix equation \cite{debaerdemacker}:
\[
\sum_{j=0}^7 c_j \left( \begin{array}{ccccc}  R^{(1)}_j &&&& \\
                                             & R^{(2)}_j &&& \\
                                             && R^{(3)}_j && \\                                            
                                             &&& R^{(4)}_j & \\ 
                                             &&&& R^{(5)}_j  \end{array} \right) =
                 \left( \begin{array}{ccccc}  U   &&&& \\
                                             & u^{(2)} &&& \\
                                             && u^{(3)} && \\                                            
                                             &&& u^{(4)} & \\ 
                                             &&&& u^{(5)}  \end{array} \right) \ , 
\]
where $u^{(2)}$, $u^{(3)}$, $u^{(4)}$, and $u^{(5)}$ are arbitrary complex numbers with unit modulus.

This equality of two $6 \times 6$ matrices
constitutes a set of eight scalar equations in the eight unknowns $c_j$:
\begin{eqnarray}
c_0 -  c_1 + c_4 - c_5 & = & U_{0,0} \nonumber \\
c_2 -  c_3 + c_6 - c_7 & = & U_{0,1} \nonumber \\ 
c_2 -  c_3 - c_6 + c_7 & = & U_{1,0} \nonumber \\
c_0 -  c_1 - c_4 + c_5 & = & U_{1,1} \nonumber \\
c_0 +  c_1 + c_2 + c_3 + c_4 + c_5 + c_6 + c_7 & = & u^{(2)} \nonumber \\
c_0 +  c_1 - c_2 - c_3 + c_4 + c_5 - c_6 - c_7 & = & u^{(3)} \nonumber \\ 
c_0 +  c_1 + c_2 + c_3 - c_4 - c_5 - c_6 - c_7 & = & u^{(4)} \nonumber \\
c_0 +  c_1 - c_2 - c_3 - c_4 - c_5 + c_6 + c_7 & = & u^{(5)} \ . \label{8vgln}
\end{eqnarray}
We find the following solution (Appendix~A):
\begin{eqnarray}
c_0 = &   & ( U_{0,0}+U_{1,1}) / 4 + (u^{(2)} + u^{(3)} + u^{(4)} + u^{(5)}) / 8 \nonumber \\
c_1 = & - & ( U_{0,0}+U_{1,1}) / 4 + (u^{(2)} + u^{(3)} + u^{(4)} + u^{(5)}) / 8 \nonumber \\
c_2 = &   & ( U_{0,1}+U_{1,0}) / 4 + (u^{(2)} - u^{(3)} + u^{(4)} - u^{(5)}) / 8 \nonumber \\
c_3 = & - & ( U_{0,1}+U_{1,0}) / 4 + (u^{(2)} - u^{(3)} + u^{(4)} - u^{(5)}) / 8 \nonumber \\
c_4 = &   & ( U_{0,0}-U_{1,1}) / 4 + (u^{(2)} + u^{(3)} - u^{(4)} - u^{(5)}) / 8 \nonumber \\
c_5 = & - & ( U_{0,0}-U_{1,1}) / 4 + (u^{(2)} + u^{(3)} - u^{(4)} - u^{(5)}) / 8 \nonumber \\
c_6 = &   & ( U_{0,1}-U_{1,0}) / 4 + (u^{(2)} - u^{(3)} - u^{(4)} + u^{(5)}) / 8 \nonumber \\
c_7 = & - & ( U_{0,1}-U_{1,0}) / 4 + (u^{(2)} - u^{(3)} - u^{(4)} + u^{(5)}) / 8 \ . \label{solu}
\end{eqnarray}
One easily checks that $\sum_{j=0}^7 |c_j|^2 = 1$. See Appendix~A.
By choosing $u^{(2)}=1$, we additionally guarantee that $\sum_{j=0}^7 c_j = 1$.
If, moreover, we also choose $u^{(3)}=u^{(4)}=u^{(5)}=1$, then we have a compact
expression for the eight weights \cite{debaerdemacker}:
\[
c_j = \delta_{j,0} + \frac{1}{4}\ \mbox{Tr}(R_j^{(1)}U) - \frac{1}{4}\ \mbox{Tr}(R_j^{(1)}) \ .
\]
Here, $\mbox{Tr}(R_j^{(1)})$ is the character $\chi_j^{(1)}$, equal to~0,
except $\chi_0^{(1)}=2$ and $\chi_1^{(1)}=-2$.

One might observe that $M_1=-M_0$, $M_3=-M_2$, $M_5=-M_4$, and $M_7=-M_6$,
such that the sum $c_0M_0 + c_1M_1 + c_2M_2 + c_3M_3 + c_4M_4 + c_5M_5 + c_6M_6 + c_7M_7$
leads to a second decomposition (in terms of the projective group): 
\begin{equation}
(c_0-c_1)M_0 + (c_2-c_3)M_2 + (c_4-c_5)M_4 + (c_6-c_7)M_6\ .
\label{c0-c1}
\end{equation}
Appendix~A demonstrates that the sum of the squares of the moduli 
of the four coefficients equals unity. 
However, we cannot guarantee that the sum of the coefficients, i.e.\ 
$(c_0-c_1) + (c_2-c_3) + (c_4-c_5) + (c_6-c_7)$, equals unity, 
as eqns (\ref{-}) impose that this sum is equal to $U_{0,0} + U_{0,1}$,
independent of the values we choose for the parameters
$u^{(2)}$, $u^{(3)}$, $u^{(4)}$, and $u^{(5)}$.

\section{The groups DS($2^w$)}

In the present section, we apply the dihedral group to quantum circuits. 
\begin{definisie}
A single-qubit circuit represented by one of the eight matrices of the group {\bf D}
is called a {\tt D} gate.  
\end{definisie}
\begin{definisie}
A $w$-qubit circuit consisting of a single {\tt D} gate on each of the $w$ wires
is called a {\tt D} stack.  
\end{definisie}
The group DS($2^w$) consists of all possible {\tt D} stacks and hence
an element of the group is represented by the Kronecker product
\begin{equation}
D_0 \otimes D_1 \otimes ... \otimes D_{w-1} \ ,
\label{kron}
\end{equation}
where each $D_j$ is a member of the group {\bf D}.

\begin{definisie}
A {\tt D} stack with all {\tt D} gates either an {\tt I} gate or an   {\tt X} gate
is called an   {\tt X} stack;    
a~{\tt D} stack with all {\tt D} gates either an {\tt I} gate or a    {\tt Z} gate
is called a    {\tt Z} stack;    
a~{\tt D} stack with all {\tt D} gates either an {\tt I} gate or a $-${\tt I} gate
is called a $-${\tt I} stack.    
\end{definisie}

\begin{lemma}
Any {\tt D} stack can be synthesised by a cascade of 
one {\tt Z} stack, one {\tt X} stack, and one {$-$\tt I} stack.
\label{ZXI}
\end{lemma}
To prove this, it suffices to observe that each {\tt D} gate of the stack 
can be decomposed as follows:
\bea
M_0 & = & {\tt Z}^0\, {\tt X}^0\, (-{\tt I})^0 \\
M_1 & = & {\tt Z}^0\, {\tt X}^0\, (-{\tt I})^1 \\
M_2 & = & {\tt Z}^0\, {\tt X}^1\, (-{\tt I})^0 \\
M_3 & = & {\tt Z}^0\, {\tt X}^1\, (-{\tt I})^1 \\
M_4 & = & {\tt Z}^1\, {\tt X}^0\, (-{\tt I})^0 \\
M_5 & = & {\tt Z}^1\, {\tt X}^0\, (-{\tt I})^1 \\
M_6 & = & {\tt Z}^1\, {\tt X}^1\, (-{\tt I})^0 \\
M_7 & = & {\tt Z}^1\, {\tt X}^1\, (-{\tt I})^1 \ .
\eea
We note that the three exponents together form 
the binary notation of the subscript~$j$ of the matrix~$M_j$. 
The sequence ({\tt Z}, {\tt X}, $-${\tt I}) is called a
transversal \cite{klap} of the group {\bf D}.
In (\ref{kron}), each $D_j$ is a $2 \times 2$ matrix from the group~{\bf D}
and thus we have:
\[
D_j = {\tt Z}^{b_j}  {\tt X}^{a_j} (-{\tt I})^{d_j} \ ,  
\]
where $b_j \in \{ 0, 1\}$,  $a_j \in \{ 0, 1\}$, and $d_j \in \{ 0, 1\}$.
We introduce the column vectors
${\bf b} = (b_0, b_1, ..., b_{w-1})^T$ and 
${\bf a} = (a_0, a_1, ..., a_{w-1})^T$.   

\begin{lemma}
The {\tt Z} stacks form a group of order~$2^w$.
\label{Z}
\end{lemma}
The group consists of $2^w \times 2^w$ diagonal matrices~$\zeta$.
Its diagonal entries~$\zeta_{k,k}$ are equal to $(-1)^{{\bf b}^T {\bf .} {\bf k}}$,
where {\bf k} denotes the column vector $(k_0, k_1, ..., k_{w-1})^T$
of the binary representation of the number~$k$.
For all matrices~$\zeta$, 
we have that the upper-left entry $\zeta_{0,0}$ equals~1.  
If ${\bf b}    = 0$, then $\zeta$ is the $2^w \times 2^w$ unit matrix~$J$.
If ${\bf b} \neq 0$, then half of the diagonal entries of $\zeta$ are equal to~$1$,
the other half being equal to~$-1$.
Indeed: let $b_p$ be the least-significant non-zero bit of ${\bf b}$.
Then the two diagonal entries $\zeta_{k,k}$ and $\zeta_{k',k'}$
(with $k$ and $k'$ equal numbers except for the bits $k_p$ and $k'_p$)
will be different, one being equal to~$1$, the other to~$-1$.
It is clear that we have $2^{w-1}$ such pairs $(k, k')$
on the diagonal of~$\zeta$. 

The group of {\tt Z} stacks is isomorphic to the direct product~{\bf C}$_2^w$.
We denote it by ZS($2^w$).
If $w>1$, then all the members of the group have determinant equal to~1.

\begin{lemma}
The {\tt X} stacks form a group of order~$2^w$.
\label{X}
\end{lemma}
The group consists of $2^w \times 2^w$ permutation matrices~$\chi$ with
entries $\chi_{k,l}$ equal to $\delta_{{\bf l},{\bf a}+{\bf k}}$, 
where {\bf l} denotes the vector $(l_0, l_1, ..., l_{w-1})^T$ and
where the sum is a bitwise addition modulo~2.
We denote the group by XS($2^w$).
It is isomorphic to the direct product~{\bf C}$_2^w$.
This is no surprise, realizing that 
it is isomorphic to ZS($2^w$), because we have $H{\tt Z}H = {\tt X}$,
where $H$ is the Hadamard matrix. 
If $w>1$, then all the members of the group have determinant equal to~1.

\begin{lemma}
The $-${\tt I} stacks form a group of order~$2$.
\label{I}
\end{lemma}
The group consists of the $2^w \times 2^w$ unit matrix~$J$ and the $2^w \times 2^w$
diagonal matrix with all diagonal entries equal to~$-1$, i.e.\ matrix~$-J$.
The non-zero entries thus equal $(-1)^d$, where $d = (d_0+d_1+ ... + d_{w-1})~\mbox{mod}~2$.
Both members of the group have determinant equal to~1.
The group is isomorphic to the cyclic group~{\bf C}$_2$.
\\

Because of Lemma~\ref{ZXI},
a {\tt D} stack is a cascade of a {\tt Z}~stack, an {\tt X}~stack, and a $-${\tt I}~stack.
Hence, each member of DS($2^w$) is a product of a diagonal matrix,
a permutation matrix and a $\pm 1$~scalar.
All $2^w \times 2^w \times 2$ products yield distinct matrices. Thus: 

\begin{lemma} 
The {\tt D} stacks form a group DS($2^w$) of order $2 \times 4^w$. 
\label{K1}
\end{lemma}
These matrices are signed permutation matrices.
The fact that the $-${\tt I}~stacks reduce to a group of order two,
allows us to reduce the $-${\tt I}~stack in Lemma~\ref{ZXI} to just one $-${\tt I}~gate.
This $-${\tt I} gate may be located on any of the $w$~wires of the circuit.

\begin{lemma} 
The group DS($2^w$) consists of $4^w$ couples
                $\{ S_{2j}, S_{2j+1}\}$, such that $S_{2j+1}=-S_{2j}$. 
\label{K2}
\end{lemma}
Indeed: if $S$ is a member of DS($2^w$), then, because $-J$ is also a member of DS($2^w$),
we have that $(-J)S = -S$ belongs to DS($2^w$).
The $4^w$~matrices $S_{2j}$ constitute a projective group.

\begin{lemma} 
The group DS($2^w$) consists of signed permutation matrices: 
                \begin{itemize}
                \item $2^w$ matrices with all $2^w$ non-zero entries equal to~$1$; 
                \item $2^w$ matrices with all $2^w$ non-zero entries equal to~$-1$;
                \item $2 \times 2^w(2^w-1)$ matrices with $2^w/2$ entries equal to~$1$
                                                      and $2^w/2$ entries equal to~$-1$.
                \end{itemize} 
\label{K3}
\end{lemma}

The group DS($2^w$) is isomorphic to one of the two
extraspecial 2-groups \cite{wiki} of order $2^{2w+1}$
(i.e.\ the one of `type +'), denoted {\bf E}$_{2^{2w+1}}^+$.
This group \cite{nebe} is a subgroup of the Pauli group \cite{gottesman} \cite{planat},
which has order $2^{2w+2}$. 

We note that, as soon as one of the factors $D_j$ of the Kronecker product (\ref{kron})
is not diagonal 
(i.e.\ as soon as one of the factors belongs to the set $\{ M_2, M_3, M_6, M_7\}$),
all diagonal entries of the product are equal to~0.
In contrast, if all factors are diagonal
(i.e.\ if all factors belong to the set $\{ M_0, M_1, M_4, M_5\}$),
then all diagonal entries of the product are equal to~$\pm 1$. Moreover, 
half of the diagonal entries equals~$1$ and 
half of the diagonal entries equals~$-1$,
except for two cases: 
 $J$ has all diagonal entries equal to~$1$ and 
$-J$ has all diagonal entries equal to~$-1$.
We conclude that all members of the group DS($2^w$) are traceless,
except for $\mbox{Tr}(J)=2^w$ and $\mbox{Tr}(-J)=-2^w$.
Hence
\[
   \sum_j                        |\mbox{Tr}(S_j)|^2 = |\, 2^w\, |^2 + |\, -2^w\, |^2 = 2 \times 4^w \ .
\]
This demonstrates the fact that the $2^w \times 2^w$ signed permutation matrices
representing the DS($2^w$) circuits
form an irreducible representation of {\bf E}$_{2^{2w+1}}^+$.
Indeed, the irreducibility criterion
\[
   \sum_j                        |\mbox{Tr}(S_j)|^2 = \mbox{Order}({\bf E}_{2^{2w+1}}^+) 
\]
is fulfilled.

An arbitrary member $S_j$ of the group DS($2^w$) 
has three parameters:
\begin{itemize}
\item the vector {\bf b},
\item the vector {\bf a}, and
\item the scalar $d$.
\end{itemize}
This means that the subscript~$j$ is a short-hand notation for
$({\bf b}, {\bf a}, d)$.  
The even subscripts~$j$ are used for matrices with $d=0$ and
the  odd subscripts~$j$ are used for matrices with $d=1$. 
The entries of the matrix $S_j$ are
\[
(S_j)_{k,l} = (-1)^{d + {\bf b}^T {\bf .} {\bf k}}\ \delta_{{\bf l},{\bf a}+{\bf k}} \ ,
\]
where      the components of the vectors {\bf a} and {\bf k} are bitwise added modulo~2.
For $w>1$, the matrix $S_j$ has unit determinant. 
Indeed, above we have seen that, for $w>1$, 
any {\tt Z} stack, any {\tt X} stack, and any $-${\tt I} stack
are represented by a matrix with determinant equal to~1. 

If    {\tt Z}$_j$ is the    {\tt Z} gate acting on the $j$th qubit,
if    {\tt X}$_j$ is the    {\tt X} gate acting on the $j$th qubit, and
if $-${\tt I}$_0$ is the $-${\tt I} gate acting on the $0$th qubit, then
the sequence ({\tt Z}$_0$, {\tt Z}$_1$, ..., {\tt Z}$_{w-1}$,
              {\tt X}$_0$, {\tt X}$_1$, ..., {\tt X}$_{w-1}$, $-${\tt I}$_0$)
is a transversal of DS($2^w$).
The sequence ($b_0, b_1, ..., b_{w-1}, a_0, a_1, ..., a_{w-1}, d\, $) is a binary number
addressing unambiguously a particular member of DS($2^w$).
In the following, we will present 
two decompositions of a unitary matrix~$U$ using the dihedral group, 
one (Section~4) where the sum of the weights will not necessarily equal~1, 
and a second, related, one (Section~5) where the sum of the weights is
constrained to~1.

\section{First decomposition of the unitary matrix}

The $2^{2w+1}$ matrices $S_j$ of the group DS($2^w$) are linearly dependent, 
as e.g.\ we have $S_0 + S_1 = J + (-J) = 0$. 
In contrast, we have 
\begin{lemma}
The $2^{2w}$ matrices $S_{2j}$ of the projective group are linearly independent.
\end{lemma}
We prove this by contradiction.
Indeed, assume that a list $(\alfa_0, \alfa_2, \alfa_4, ...,$ $\alfa_{2^{2w+1}-2})$,
different from the zero list $(0, 0, 0, ..., 0)$, exists, such that
\[
\sum_{j=0}^{2^{2w}-1} \alfa_{2j}S_{2j} = 0 \ .
\]
We multiply both sides of this equation to the left with $S_{2k}^T$,
where $k$ is any integer from $(0, 1, 2, ..., 2^{2w}-1)$.
Subsequently, we take the trace of both sides of the equation.
According to Appendix~B, 
we find $\alfa_{2k} 2^{2w} = 0$ for all~$k$, and thus all $\alfa_{2k}=0$.
Hence, the list $(\alfa_0, \alfa_2, \alfa_4, ..., \alfa_{2^{2w+1}-2})$ is the zero list,
in contradiction with the assumption.
This proof is reminiscent of the proof by Veltman \cite{martinus} \cite{veltman}
of a similar property of $4 \times 4$ gamma matrices. 

Because the $2^{2w}$ matrices $S_{2j}$ thus form a complete set of $2^w \times 2^w$ matrices,
we have:
\begin{teorema} 
Any U($2^w$) matrix $U$, i.e.\ any matrix representing a $w$-qubit quantum circuit,
can be written 
       \begin{equation}
       U = \sum_j g_{2j} S_{2j}
       \label{martin}
       \end{equation}
       with all $S_{2j}$ member of the projective group associated to DS($2^w$) 
       and the weights $g_{2j}$ complex numbers.
\label{3}
\end{teorema}
Multiplying (\ref{martin}) to the left by $S_{2k}^T$ and taking traces,
leads to $\mbox{Tr}(S_{2k}^TU) = g_{2k} 2^{w}$ and thus to the value of the weights:
\begin{equation}
g_{2k} = 2^{-w}\ \mbox{Tr}(S_{2k}^TU) \ .
\label{newc}
\end{equation}

According to Appendix~C, we have
\bea
\sum_j g_{2j} & = & 2^{-w}\ \sum_j \mbox{Tr}(S_{2k}^TU) = 
                    2^{-w}\ 2^w  \sum_l U_{0,l} = \sum_l U_{0,l} \ .
\eea
Thus the sum of the weights equals the uppermost row sum of the matrix~$U$,
a number not necessarily equal to~1. 
Using the short-hand notation $n=2^w$, we note that
\bea 
|g_{2j}|^2 & = & \frac{1}{n^2}\ \left| \mbox{Tr}(S_{2j}^TU)\right| ^2            \\[1mm]
           & = & 1 - \left(  1 - \frac{\ |\mbox{Tr}(S_{2j}^TU)|^2\ }{n^2}\right) \\[1mm]
           & = & 1 - D(S_{2j}, U) \ , 
\eea
where 
\[
D(A, B) = 1 - \frac{\ \left|\mbox{Tr}(A^{\dagger}B)\right|^2}{n^2}
\] 
is the distance
between the $n \times n$ unitary matrices~$A$ and~$B$, according to Khatri et al. \cite{khatri},
the trace $\mbox{Tr}(A^{\dagger}B)$ being known as
the Hilbert--Schmidt inner product of~$A$ and~$B$.
Hence, the nearer the $S_{2j}$~matrix is to the given matrix~$U$,
the more it contributes to the decomposition of~$U$. 
Finally, we have 
\[
\sum_j |g_{2j}|^2 = 1 \ . 
\]
Proof is in the Appendix~C.

As an example, we decompose the unitary transformation
\begin{equation}
U = \frac{1}{12}\
\left( \begin{array}{cccc} 8    &  4+8i & 0    &  0    \\
                           2+ i &   -2i & 3-9i & -3-6i \\
                           1-7i & -6+2i & 6    & -3+3i \\
                           3+4i &  2-4i & 3-3i &    9i \end{array} \right) \ .
\label{onzeboek blz. 75}
\end{equation}
Its decomposition according to Theorem~\ref{3} and eqn (\ref{newc}) is
\bea
&& g_0S_0 + g_2S_2 + g_4S_4 + ... + g_{30}S_{30} = \\[1mm] && 
\frac{14 + 7i}{48}\ \left( \begin{array}{cccc}  1 &&&    \\ &   1 && \\ &&  1 & \\   &&&  1  \end{array} \right) + \   
\frac{ 2 -11i}{48}\ \left( \begin{array}{cccc}  1 &&&    \\ &   1 && \\ && -1 & \\   &&& -1  \end{array} \right) \\[1mm] &+& 
\frac{14 - 7i}{48}\ \left( \begin{array}{cccc}  1 &&&    \\ &  -1 && \\ &&  1 & \\   &&& -1  \end{array} \right) + \ ... \ +  
\frac{ 6 +11i}{48}\ \left( \begin{array}{cccc}    &&&  1 \\ && -1 &  \\ & -1 && \\ 1 &&&     \end{array} \right) \ . 
\eea
In this particular example, the 16~distances $D(U, S_{2j})$ vary from
$2015/2304 \approx 0.876$ to
$2291/2304 \approx 0.994$.

\section{Second decomposition of the unitary matrix}

After Klappenecker and R\"otteler \cite{klap} 
and De Baerdemacker et al.\ \cite{debaerdemacker}
and taking into account that any finite group has
the trivial 1-dimensional irreducible representation, we have
\begin{teorema}
If a unitary matrix $U$ can be written 
       \[
       U = \sum_j g_j G_j
       \]
       with all $G_j$ member of some finite group~{\bf G}, 
       then there exists a decomposition
       \[
       U = \sum_j h_j G_j \ ,
       \]
       such that both $\sum_j  h_j    = 1$ 
       and            $\sum_j |h_j|^2 = 1$.
\label{8}
\end{teorema}
Together, Theorems~\ref{3} and~\ref{8} lead to the final result:  
\begin{teorema}
Any U($2^w$) matrix $U$, i.e.\ any matrix representing a $w$-qubit quantum circuit,
can be written 
       \begin{equation}
       U = \sum_j h_j S_j
       \label{cS}
       \end{equation}
       with all $S_j \in$ DS($2^w$) 
       and the weights $h_j$ complex numbers,
       such that both $\sum_j h_j = 1$ and $\sum_j |h_j|^2 = 1$.
\label{teo4}
\end{teorema}

From \cite{debaerdemacker} we have a closed form for the weights appearing in (\ref{cS}):
\begin{eqnarray}
h_j = \frac{1}{N} \sum_{\nu=1}^{\mu} n_{\nu}\ \mbox{Tr}
                 \left( R^{(\nu) \, \dagger}_j\, U^{(\nu)}_j\right) \ , 
\label{formule}
\end{eqnarray}
where $\mu$ is the number of irreducible representations of $S_j$, 
where $n_{\nu}$ is the dimension of the particular irrep $R^{(\nu)}_j$, and
where $N$ is the order of the group~{\bf G}.

If, for $U^{(1)}_j$ we choose the given matrix~$U$ and
for each matrix $U^{(\nu)}_j$ with $2 \le \nu \le \mu$
we choose the $n_{\nu} \times n_{\nu}$ unit matrix,
then (\ref{formule}) becomes
\begin{equation}
h_j = \frac{1}{N}\ 
             \left[\                  n_1    \, \mbox{Tr}\left(R^{(1)  \, \dagger}_j\, U \right) + 
              \sum_{\nu=2}^{\mu} n_{\nu}\, \mbox{Tr}\left(R^{(\nu)\, \dagger}_j     \right) \ \right] \ .
\label{formule1}
\end{equation} 
We take advantage of Schur's orthogonality relation:
\[
 \sum_{\nu} n_{\nu}\,  \mbox{Tr}\left(R^{(\nu)\, \dagger}_j\right) = 
 \sum_{\nu} n_{\nu}\,  \mbox{Tr}\left(R^{(\nu)\, \dagger}_j\, R^{(\nu)}_0\right) =
 \delta_{0,j}\, N \ .
\]
Because moreover $n_1=2^w$ and $N=2^{2w+1}$, 
we obtain the explicit expression for the weight:
\bea
h_j & = & \delta_{0,j}+ 
             \frac{1}{2^{w+1}}\ \mbox{Tr}\left(R^{(1)\, \dagger}_j\, U \right)  - 
             \frac{1}{2^{w+1}}\ \mbox{Tr}\left(R^{(1)\, \dagger}_j\,   \right)     \\
    & = & \delta_{0,j}+ 
             \frac{1}{2^{w+1}}\ \mbox{Tr}(S_j^T\, U)  - 
             \frac{1}{2^{w+1}}\ \mbox{Tr}(S_j)\ .  
\eea
As demonstrated in Appendix~B, we have $\mbox{Tr}(S_0)=2^w$, $\mbox{Tr}(S_1)=-2^w$,
and $\mbox{Tr}(S_j)=0$ for $j>1$. Hence:
\begin{eqnarray}
h_0 & = &      \frac{1}{2^{w+1}}\ \mbox{Tr}(U) + \frac{1}{2} \nonumber \\
h_1 & = & - \, \frac{1}{2^{w+1}}\ \mbox{Tr}(U) + \frac{1}{2} \nonumber \\
h_j & = &      \frac{1}{2^{w+1}}\ \mbox{Tr}(S_j^T\, U) \hspace{10mm} \mbox{ for } j > 1 \ .
\label{chi} 
\end{eqnarray} 

Matrix example (\ref{onzeboek blz. 75}),
according to Theorem~\ref{teo4} and eqn (\ref{chi}), has decomposition
\bea
&& h_0S_0 + h_1S_1 + h_2S_2 + ... + h_{31}S_{31} = \\[1mm] && 
\frac{62 + 7i}{96}\ \left( \begin{array}{cccc}  1 &&&    \\ &   1 && \\ &&  1 & \\    &&&  1  \end{array} \right) + \   
\frac{34 - 7i}{96}\ \left( \begin{array}{cccc} -1 &&&    \\ &  -1 && \\ && -1 & \\    &&& -1  \end{array} \right) \\[1mm] &+& 
\frac{ 2 -11i}{96}\ \left( \begin{array}{cccc}  1 &&&    \\ &   1 && \\ && -1 & \\    &&& -1  \end{array} \right) + \ ... \ +  
\frac{-6 -11i}{96}\ \left( \begin{array}{cccc}    &&& -1 \\ &&  1 &  \\ &  1 && \\ -1 &&&     \end{array} \right) \ .
\eea

\section{Generalization}

The above conclusions for    arbitrary U($2^w$) matrices 
can easily be generalized to arbitrary U($p^w$) matrices,
where $p$ is an arbitrary prime. 
Indeed, let $\w$ be the $p\,$th root of~1.
We define the X gate and Z gate by their respective $p \times p$ matrices:
\[
X = \left( \begin{array}{ccccc}      & 1      &&&   \\
                                     && 1      &&   \\

                                     &&& \ddots &   \\
                                     &&&&         1 \\
                                   1 &&&&           \end{array} \right)
\hspace{4mm} \mbox{ and } \hspace{4mm}
Z = \left( \begin{array}{cccccc}    1            &&&&&          \\

                                   & \w           &&&&          \\
                                   && \ddots       &&&          \\

                                   &&&&     \w^{p-2} &          \\

                                   &&&&&               \w^{p-1} \end{array} \right) \ .
\]
We have
\[ 
X^p = Z^p = I \ ,
\]
where $I$ is the $p \times p$ unit matrix. Moreover, we have
\[
XZ = \w ZX \ .
\]
As a result, any matrix generated by the two generators $X$ and $Z$
can be written as $Z^bX^a\w^d$ with 
$b \in \{ 0, 1, 2, ..., p-1\}$,
$a \in \{ 0, 1, 2, ..., p-1\}$, and
$d \in \{ 0, 1, 2, ..., p-1\}$.
Therefore, the group generated by $X$ and $Z$ is the 
extraspecial group ${\bf E}_{p^3}^{+}$ of order $p^3$.
This group takes over the role of the dihedral group ${\bf D} = {\bf E}_{8}^{+}$.

Any gate generated by $X$ and $Z$, we call an $E$~gate.
A circuit acting on $w$ qudits and consisting of a single $E$~gate
on each of the $w$~wires, we call an $E$~stack.
The $E$~stacks form a group isomorphic to ${\bf E}_{p^{2w+1}}^{+}$ of order $p^{2w+1}$.
An arbitrary $E$ stack is represented by 
a $p^w \times p^w$ complex permutation matrix $C_j$ with entries
\[
(C_j)_{k,l} = \w^{d + {\bf b}^T {\bf .} {\bf k}}\ \delta_{{\bf l},{\bf a}+{\bf k}} \ ,
\]
where $+$ stands for addition modulo~$p$.
The Hilbert--Schmidt inner product $T_{j,k}=\mbox{Tr}(C_j^TC_k)$
of two such matrices equals $\w^qp^w$ if $C_k=\w^qC_j$ 
for some $q \in \{ 0, 1, 2, ..., p-1 \}$ and equals zero otherwise.
This fact leads to a decomposition of an arbitrary U($p^w$) matrix:
\[
U = \sum_j g_{pj}C_{pj} \ ,
\]
with all $C_{pj}$ member of the projective group of order $p^{2w}$, associated to
${\bf E}_{p^{2w+1}}^{+}$,
and with the weights $g_{pj}$ being equal to $p^{-w}\,\mbox{Tr}(C_{pj}^TU)$
and having the property $\sum_j |g_{pj}|^2=1$.
Finally, this leads to a second decomposition:
\[
U = \sum_j h_{j}C_{j} \ ,
\]
with all $C_{j}$ member of the group
${\bf E}_{p^{2w+1}}^{+}$ of order $p^{2w+1}$
and with the weights $h_{j}$
having the two properties $\sum_j |h_j|^2=1$ and $\sum_j h_j=1$.

\section{Conclusion}

We conclude that a unitary matrix, describing an arbitrary $w$-qubit quantum circuit,
i.e.\ a member of the matrix group U($2^w$),
can be decomposed as a weighted sum of a finite number of signed permutation matrices,
each describing a stack of $w$~gates, 
each a single-qubit dihedral gate.
The weights of the sum add up to~1, just like the squares of the moduli of these weights.
The signed permutation matrices belong to a subgroup
isomorphic to the extraspecial group {\bf E}$_{2^{2w+1}}^+$.
The order of this group is $2^{2w+1}$.
A projective group of order $2^{2w}$ suffices for the decomposition
if we do not impose that the sum of the weights is equal to~1,
i.e.\ if we only impose that the sum of squared moduli of the weights equals unity.
Similar conclusions hold for members of the unitary matrix group U($p^w$),
with $p$ an arbitrary prime.

\section*{Acknowledgement} 

This research was undertaken, in part, 
thanks to funding from the Canada Research Chairs program.

\clearpage

\appendix

\section{Detailed calculations for U(2)}

The former four equations of (\ref{8vgln}) yield
\begin{eqnarray}
c_0 - c_1 & = & (U_{0,0} + U_{1,1}) / 2 \nonumber \\
c_2 - c_3 & = & (U_{0,1} + U_{1,0}) / 2 \nonumber \\
c_4 - c_5 & = & (U_{0,0} - U_{1,1}) / 2 \nonumber \\
c_6 - c_7 & = & (U_{0,1} - U_{1,0}) / 2 \ ; \label{-}
\end{eqnarray}
the latter four equations yield
\begin{eqnarray}
c_0 + c_1 & = & (u^{(2)} + u^{(3)} + u^{(4)} + u^{(5)}) / 4 \nonumber \\
c_2 + c_3 & = & (u^{(2)} - u^{(3)} + u^{(4)} - u^{(5)}) / 4 \nonumber \\
c_4 + c_5 & = & (u^{(2)} + u^{(3)} - u^{(4)} - u^{(5)}) / 4 \nonumber \\
c_6 + c_7 & = & (u^{(2)} - u^{(3)} - u^{(4)} + u^{(5)}) / 4 \ . \label{+}
\end{eqnarray}
These results immediately lead to the solution (\ref{solu}).

Additionally, the four eqns (\ref{-}) lead to
\[
|c_0-c_1|^2 + |c_2-c_3|^2 + |c_4-c_5|^2 + |c_6-c_7|^2 =  
\frac{1}{2}\ (\ |U_{0,0}|^2  + |U_{0,1}|^2  + |U_{1,0}|^2  + |U_{1,1}|^2  \ ) = 1 \ ,
\]
while the four eqns (\ref{+}) lead to
\[
|c_0+c_1|^2 + |c_2+c_3|^2 + |c_4+c_5|^2 + |c_6+c_7|^2 =  
\frac{1}{4}\ (\ |u^{(2)}|^2 + |u^{(3)}|^2 + |u^{(4)}|^2 + |u^{(5)}|^2 \ ) = 1 \ .
\]
The identities 
\bea
|c_0|^2 + |c_1|^2 & = & |c_0-c_1|^2 / 2 \ +\  |c_0+c_1|^2 / 2 \\
|c_2|^2 + |c_3|^2 & = & |c_2-c_3|^2 / 2 \ +\  |c_2+c_3|^2 / 2 \\
|c_4|^2 + |c_5|^2 & = & |c_4-c_5|^2 / 2 \ +\  |c_4+c_5|^2 / 2 \\
|c_6|^2 + |c_7|^2 & = & |c_6-c_7|^2 / 2 \ +\  |c_6+c_7|^2 / 2   
\eea
thus yield
\[
\sum_{j=0}^7 |c_j|^2 = 1/2 \ + \ 1/2 = 1 \ .
\]

\section{Trace of signed permutation matrix}

We compute the Hilbert--Schmidt inner product $T_{j,k}$ of two
signed permutation matrices $S_j$ and $S_k$:
\bea
T_{j,k} & = & 
\mbox{Tr}(S_j^TS_k) = \sum_u (S_j^TS_k)_{u,u} =
\sum_u  \sum_p \ (S_j^T)_{u,p}(S_k)_{p,u} \\ 
& = & \sum_u  \sum_p \  (S_j)_{p,u}(S_k)_{p,u}   \\
& = & \sum_u  \sum_p \ (-1)^{d_j+{\bf b}_j^T.{\bf p}} 
                     \ \delta_{{\bf u},{\bf a}_j + {\bf p}}\ 
                       (-1)^{d_k+{\bf b}_k^T.{\bf p}}  
                     \ \delta_{{\bf u},{\bf a}_k + {\bf p}} \\
 & = & \sum_p \ (-1)^{d_j+d_k+ ({\bf b}_j^T + {\bf b}_k^T).{\bf p}} 
       \sum_u \ \delta_{{\bf u},{\bf a}_j + {\bf p}}\,  
                \delta_{{\bf u},{\bf a}_k + {\bf p}} \ .
\eea
If the eqns
\bea
{\bf u} & = & {\bf a}_j + {\bf p} \\
{\bf u} & = & {\bf a}_k + {\bf p}
\eea
are fulfilled, then the corresponding number~$u$ points to 
a $\pm 1$ entry in position $(u,u)$ of the matrix~$S_j^TS_k$.
A necessary condition for a solution is ${\bf a}_j = {\bf a}_k$.
Therefore $T_{j,k}=0$ if ${\bf a}_j \neq {\bf a}_k$.
If instead ${\bf a}_j = {\bf a}_k$, then
\[
\sum_u \ \delta_{{\bf u},{\bf a}_j + {\bf p}}\,  
                \delta_{{\bf u},{\bf a}_k + {\bf p}} = 1 
\]
and
\[ 
T_{j,k} = (-1)^{d_j+d_k} \  \sum_p \ (-1)^{ ({\bf b}_j^T + {\bf b}_k^T) . {\bf p} } \ . 
\]
If ${\bf b}_j = {\bf b}_k$, then $\sum_p \ (-1)^{ ({\bf b}_j^T + {\bf b}_k^T) . {\bf p} } = 2^w$.
If instead  ${\bf b}_j \neq {\bf b}_k$, then  $\sum_p \ (-1)^{ ({\bf b}_j^T + {\bf b}_k^T) . {\bf p} } = 0$.
Thus, iff both ${\bf a}_j = {\bf a}_k$ and ${\bf b}_j = {\bf b}_k$, then $T_{j,k} = \pm 2^w$. 
We conclude: 
\begin{itemize}
\item $T_{j,k} =  2^w$ if $S_k= S_j$,
\item $T_{j,k} = -2^w$ if $S_k=-S_j$, and
\item else $T_{j,k}=0$.
\end{itemize}

In order to calculate the trace of an arbitrary $S_k$,
it suffices to apply the above results with $S_j$ equal to $S_0=J$,
the $n \times n$ unit matrix: Tr$(S_k) = T_{0,k}$. 
Thus: Tr$(S_0)=2^w$ and Tr$(S_1)=-2^w$; all other Tr$(S_j)=0$. 

\section{Detailed calculations for U($2^w$)}

Confronting the two decompositions of $U$ (Sections~4 and~5, respectively)
reveals the following relationship between 
the $2^{2w}$   weights~$g_{2j}$ and
the $2^{2w+1}$ weights~$h_j$\,~:
\bea
g_0    & = & 2h_0 - 1 \\
g_{2j} & = & 2h_{2j}  \ \ \ \ \ \mbox{ for } j>0 \ .
\eea
Together with $h_{1}=-h_{0}+1$ and $h_{2j+1}=-h_{2j}$ for $j>0$,
this yields the inverse relationship:
\bea
h_0      & = & (1+g_0)/2                                       \\
h_1      & = & (1-g_0)/2                                       \\
h_{2j}   & = &    g_{2j}/2  \ \ \ \ \ \ \ \ \ \mbox{ for } j>0 \\
h_{2j+1} & = &   -g_{2j}/2      \ \ \ \ \ \ \ \mbox{ for } j>0 \ .
\eea
This allows us to compute the sum $\sum |g_j|^2$ from the known $\sum |h_j|^2=1\, $:
\bea
1 = \sum |h_j|^2 & = & |h_0|^2 + |h_1|^2 + \sum_{j > 0}|h_{2j}|^2  + \sum_{j > 0}|h_{2j+1}|^2     \\
                 & = & \frac{1}{4}\, |1+g_0|^2 + \frac{1}{4}\, |1-g_0|^2 + 
                       \frac{1}{4}\,  \sum_{j > 0}|g_j|^2  +
                       \frac{1}{4}\,  \sum_{j > 0}|g_j|^2     \\
                 & = & \frac{1}{2} (1+ |g_0|^2) + \frac{1}{2} \sum_{j > 0}|g_{2j}|^2 \ .
\eea
Hence:
\[
|g_0|^2 + \sum_{j > 0} |g_{2j}|^2 = 1
\]
and thus $\sum_j |g_{2j}|^2 = 1$.

For the sum of the weights~$g_j$, we compute
\bea
\sum_j\ \mbox{Tr}(S_{2j}^T\, U) & = & \sum_{k,\, l}\  \sum_j\ (S_{2j})_{k,l}\, U_{k,l} \\
& = & \sum_{k,\, l}\ U_{k,l}\,  \sum_j\ (-1)^{0+{\bf b}^T.{\bf k}}\, \delta_{{\bf l},\, {\bf a}+{\bf k}} \ .
\eea
Whereas until here $\sum_j$ means summing over the paramaters {\bf a} and {\bf b},
from now on, we can restrict the value of {\bf a} to ${\bf l}-{\bf k}$.
Thus summing only happens over the parameter {\bf b}\, :
\bea
\sum_j\ \mbox{Tr}(S_{2j}^T\, U) & = &  \sum_{k,\, l}\  U_{k,l}\ \sum_{\bf b} (-1)^{{\bf b}^T.{\bf k}}  \\
& = &   \sum_{k \neq 0,\, l}\  U_{k,l}\ \sum_{\bf b} (-1)^{{\bf b}^T.{\bf k}} 
      + \sum_{            l}\  U_{0,l}\ \sum_{\bf b} (-1)^{{\bf b}^T.{\bf 0}} \\
& = &   \sum_{k \neq 0,\, l}\  U_{k,l}\  0 + \sum_{l}\  U_{0,l}\ \sum_{\bf b} 1 \\
& = &   \sum_{l}\  U_{0,l}\ 2^w =  2^w \sum_{l}\  U_{0,l}\ .   
\eea
 So, finally, (\ref{newc}) becomes
\[
\sum_j g_{2j} = \frac{1}{2^w}\  2^w \sum_{l}\  U_{0,l} = \sum_{l}\  U_{0,l} \ .  
\] 

\end{document}